\documentclass[fleqn,usenatbib]{mnras}

\usepackage{newtxtext,newtxmath}

\usepackage[T1]{fontenc}

\DeclareRobustCommand{\VAN}[3]{#2}
\let\VANthebibliography\thebibliography
\def\thebibliography{\DeclareRobustCommand{\VAN}[3]{##3}\VANthebibliography}

\usepackage{graphicx}	

\usepackage{amsmath,amsfonts} 
\usepackage{algorithmic}
\usepackage{array}

\usepackage{multicol}
\usepackage{caption}
\usepackage{subcaption}
\usepackage{rotating}
\usepackage{wrapfig}
\usepackage{booktabs}
\usepackage{multirow}
\usepackage{adjustbox}
\usepackage{hyperref}
\usepackage{makecell}
\usepackage{url}

\usepackage{scrextend}

\usepackage{float}
\restylefloat{table}
\usepackage{fixltx2e}
\usepackage{stfloats}

\usepackage{xcolor}
\usepackage{soul}

\sethlcolor{white} 

\usepackage{lineno}

\title[CzSL: Image classification with citizen science]{CzSL: Learning from citizen science, experts and unlabelled data in astronomical image classification}

\author[Manuel Jim\'enez et al.]{
Manuel Jim\'enez,$^{1,2}$\thanks{E-mail: majm8609@gmail.com (MJ)}
Emilio J. Alfaro,$^{1}$
Mercedes Torres Torres$^{3}$ 
and Isaac Triguero$^{4,5,2}$
\\
$^{1}$Instituto de Astrof\'isica de Andaluc\'ia (CSIC), Granada 18008, Spain \\
$^{2}$School of Computer Science, University of Nottingham, Nottingham NG8 1BB, United Kingdom \\
$^{3}$B-Hive Innovations Ltd, Lincoln LN6 7DJ, United Kingdom \\
$^{4}$Andalusian Research Institute in Data Science and Computational Intelligence (DaSCI), University of Granada, Granada 18071, Spain \\
$^{5}$Department of Computer Science and Artificial Intelligence, University of Granada, Granada 18071, Spain 
}

\date{Accepted XXX. Received YYY; in original form ZZZ}

\pubyear{2023}

\begin{document}
\label{firstpage}
\pagerange{\pageref{firstpage}--\pageref{lastpage}}
\maketitle

\begin{abstract}

Citizen science is gaining popularity as a valuable tool for labelling large collections of astronomical images by the general public. This is often achieved at the cost of poorer quality classifications made by amateur participants, which are usually verified by employing smaller data sets labelled by professional astronomers. Despite its success, citizen science alone will not be able to handle the classification of current and upcoming surveys. To alleviate this issue, citizen science projects have been coupled with machine learning techniques in pursuit of a more robust automated classification. However, existing approaches have neglected the fact that, apart from the data labelled by amateurs, (limited) expert knowledge of the problem is also available along with vast amounts of unlabelled data that have not yet been exploited within a unified learning framework. This paper presents an innovative learning methodology for citizen science capable of taking advantage of expert- and amateur-labelled data, featuring a transfer of labels between experts and amateurs. The proposed approach first learns from unlabelled data with a convolutional autoencoder and then exploits amateur and expert labels via the pre-training and fine-tuning of a convolutional neural network, respectively. We focus on the classification of galaxy images from the \textit{Galaxy Zoo} project, from which we test binary, multi-class, and imbalanced classification scenarios. The results demonstrate that our solution is able to improve classification performance compared to a set of baseline approaches, deploying a promising methodology for learning from different confidence levels in data labelling. 


\end{abstract}



\section{Introduction}
\label{introduction}
Citizen science (CzS) \citep{Kullenberg2016,Show2015524} has re-emerged as a type of crowdsourcing that entails involving amateur participants in scientific research to perform large-scale distributed data gathering or data processing, typically the classification of vast collections of images on the web \citep{Lintott20081179}. Of particular note has been the great expansion of such paradigm in astronomy, with the success of the \textit{Galaxy Zoo} project in the last decade \citep{Masters2019205} and many others that have followed a similar path \citep{Fischer2012,Beaumont2014}. The main motivation behind the development of these initiatives is the need to reduce the time to classify the large amounts of data generated without involving experts in the process \citep{Bonney20141436,Lamas2021266}. Despite their success and popularity, many application areas such as astronomy or geo-sciences will soon achieve image data acquisition on a scale of billions of objects \citep{Zhang201511,Sudmanns2020832}, which would take hundreds of years to classify even with the help of the largest team of citizen scientists that have participated in any single CzS project to date \citep{Walmsley20201554}. Consequently, there is a call for the development of novel machine learning (ML) \citep{Witten2005} approaches that can effectively automate the labelling of images at a higher scale \citep{Sen2022}, ideally combining the power of humans and machines in the context of CzS \citep{Beck2018476}. 

The outcome of a CzS project generally consists of a data set labelled by project participants that is substantially larger than existing expert-labelled data sets \citep{Lintott2011166}. These amateur labels may be less reliable as they were assigned by people that tend to hold a variable set of skills and motivations \citep{Herodotou2020}. Furthermore, the final classification of an object must be deduced from a set of independent (amateur) judgements, and multiple biases are often found when amateur classifications are evaluated with the aid of expert knowledge on the problem at hand \citep{Lintott20081179,Kosmala2016551}. The methods followed to merge all information derived from citizens' efforts and tackle the inherent uncertainty have proven to be decisive to obtain the most benefits from CzS data \citep{Jimenez2019479}. Nonetheless, along with the smaller sets of expert-labelled data, there are also abundant unlabelled data, which can be leveraged to improve the learning process. 

In astronomical image classification, ML approaches have made use of CzS data to train automated classifiers that are purely based on amateur-labelled images \citep{Banerji2010342,Dieleman20151441,Cheng2021507}, without dealing with the prevalent uncertainty in this sort of data and neglecting the available expert knowledge. These works have generally exploited the potential behind deep learning (DL) \citep{Goodfellow2016} with the use of convolutional neural networks (CNNs) \citep{Bengio20091, Guo201627}. Similarly, convolutional autoencoders (CAEs) \citep{Ribeiro201813} have been employed in many applications towards the unsupervised classification of images \citep{Cheng2020494}. On the other hand, semi-supervised learning (SSL) \citep{Zhu20091} methods applied to image classification focus on improving performance by considering a combination of labelled and unlabelled data, but do not account for different confidence levels within the labelled part of the training examples. Numerous SSL approaches are found in the literature, most notably those belonging to self-labelled techniques \citep{Triguero2015245} or graph-based methods \citep{Du20191440}, which have also combined their strengths with CNNs \citep{Wu20181259}. Despite all this, the CzS data framework has not been suitably leveraged from an ML perspective, taking advantage of all levels of knowledge (expert/amateur/unlabelled) in the model's learning.  

To fill that gap, this paper presents a novel learning methodology with CzS leading to a more robust astronomical image classification. Ultimately, our goal is the joint exploitation of expert-based classifications, CzS data, and the wealth of unlabelled data prevalent nowadays in astronomical surveys. We propose the Citizen Science Learning (CzSL) methodology, an innovative DL-based algorithm capable of sequentially taking advantage of expert and amateur labels in conjunction with unlabelled data. Such an approach is based on two central elements. First, the pre-training and fine-tuning of neural networks, a class of transfer learning \citep{Pan20101345} that has been thoroughly used in the adaptation of very deep networks to problems in a specific domain \citep{Marmanis2016105,Ackermann2018415} or to facilitate their use in situations where the scarcity of labelled examples makes training the network from scratch unfeasible \citep{Kim2020294,Liu2020253}. In contrast, here we have leveraged this scheme to learn from both amateur- and expert-based classifications with a CNN, which is a strategy that has already been investigated by the authors with promising results \citep{Jimenez20208}, and have extended it to include learning from unlabelled data with a CAE. In addition, we have developed a method to extend expert knowledge to amateur classifications with a multi-layer perceptron (MLP) \citep{Lecun2015436}, which directly inputs CzS data and learns the best correlation between amateur and expert labels. 

As case study, we have considered the well-known Galaxy Zoo 1 (GZ1) project \citep{Lintott2011166}, which tackled the classification of galaxy images according to their morphology \citep{Hubble1926321,Sandage2005581}. On the basis of these data, we have covered three different classification scenarios: distinguishing between elliptical and spiral classes (binary classification), differentiating the handedness of spiral arms in the last setting \citep{Longo2011224} (multi-class classification), and merger detection \citep{Darg20101043} (imbalanced classification). We have established a set of comparative approaches considering distinct combinations of data and labelled samples, including a form of self-labelled SSL. 
 
The rest of the paper is organised as follows. In Section \ref{background}, we expand on the background of CzS and review the current literature on SSL approaches and transfer learning with CNNs. Section \ref{CzSL-proposal} presents the CzSL algorithm. In Section \ref{experiments}, we explain the experiments carried out to test the capability of such an approach, and Section \ref{results-analysis} presents the results and analysis. Finally, in Section \ref{conclusions} we draw some conclusions and outline potential directions for future work.

\section{Related work}
\label{background}
This section provides a brief overview of current trends in CzS and DL concepts covered in the paper. First, we introduce in more depth the CzS approach along with the GZ1 project and current attempts to solve the galaxy classification problem with ML (Section \ref{citizen-science-data}). Then, we examine the landscape of SSL with CNNs to identify similarities with the proposed approach and review the DL-based methods used in the implementation of CzSL (Section \ref{beyond-supervised-classification}).

\subsection{The use of citizen science to boost automated galaxy classification}
\label{citizen-science-data}
CzS has been standard practice since the earliest stages of modern science \citep{Silvertown2009467}. However, the world-wide spread of Internet and the emergence of information technologies (cloud computing, geo-sensors, mobile devices, etc.) have greatly accelerated the rise of numerous CzS-based initiatives \citep{Lamas2021266}, fostering the exploitation of these resources for modern CzS in web platforms that bring real science to amateurs' homes and facilitate their development by professional researchers \citep{Bonney20141436,Newman2012298}. Nowadays, online CzS projects covering multiple research problems in a variety of disciplines \citep{Simpson20141049} provide distributed data analysis at a scale that is unfeasible for experts alone, and are having an impact on research \citep{Follett201511,Masters2019205}. 

Amongst others, classification is one of the most demanded fundamental tasks in such projects \citep{Kullenberg2016,Trouille20191902}. Typically, volunteer participants are asked to complete the classification of images that are sequentially displayed in the project website by clicking on a set of pre-defined choices. In the case of GZ1, developers also included an option for blurry/undecidable images, that is, a \textit{Don't Know} alternative \citep{Lintott2011166}. The resulting data look like a set of scores that are called ``vote fractions'', defined as the ratio of the number of votes for a particular class to the total number of votes. Assuming that they are unbiased, then they approximate the probability of an object belonging to a class. An extract of these data is shown in Table \ref{tab:citizen-science-data}. Therefore, analysis of these data is crucial to achieve correct classifications and to get the best benefit from this \textit{people-driven} research \citep{Jimenez2019479}. Previous works have already investigated different strategies to overcome this issue. For instance, many methods have been proposed to counteract the effect of biases by applying basic statistics to results \citep{Lintott2011166,Bamford20091324,Kosmala2016551} or guiding participants through the classification task during the course of the project \citep{Crowston2020123,Walmsley2022}. Expert knowledge has also been employed to assess the confidence in amateur classifications \citep{Lintott20081179,Swanson2016520} or enhance the CzS data utility \citep{Jimenez2019479}. 

\begin{table}[!h]
\centering
\resizebox{1.00\columnwidth}{!}{
\begin{tabular}{cccccc}

\hline                       
\textbf{\textit{Image ID}} & \textbf{\textit{Votes}} & \textbf{\textit{Elliptical}} & \textbf{\textit{Spiral}} & $ \dotsi $ & \textbf{\textit{Don't Know}} \\\hline

587727178449485858 & 24 & 0.125 & 0.875 & $ \dotsi $ & 0.000 \\
588015509806252152 & 38 & 0.711 & 0.132 & $ \dotsi $ & 0.158 \\
587730773351858407 & 64 & 0.625 & 0.282 & $ \dotsi $ & 0.078 \\
588015508195639450 & 42 & 0.429 & 0.357 & $ \dotsi $ & 0.214 \\
588017721180291223 & 26 & 1.000 & 0.000 & $ \dotsi $ & 0.000 \\
$ \vdots $ & $ \vdots $ & $ \vdots $ & $ \vdots $ & $ \ddots $ & $ \vdots $ \\
587727220876705961 & 22 & 0.773 & 0.045 & $ \dotsi $ & 0.182 \\\hline 
 
\end{tabular}}
\caption{\label{tab:citizen-science-data} Extract of the data recorded in GZ1 project. \textit{Don't Know} scores account for blurry/undecidable images.}
\end{table} 

We have adopted the morphological classification of galaxy images as a case study \citep{Hubble1926321}. In the simplest version, it can be regarded as a binary classification problem consisting of two main classes: elliptical and spiral. However, the multiplicity of hybrid types amongst them along with the huge variability in galaxy shape, orientation, or brightness, turns it into a very challenging classification task, even for human classifiers \citep{Lahav1995859}.

The identification of elliptical and spiral classes was the target of GZ1 \citep{Lintott20081179,Masters2019205}, which ended up recruiting more than 200 000 participants that produced the largest manually-classified galaxy catalogue to date, including nearly 900 000 examples \citep{Lintott2011166}. In GZ1, amateur votes were spread across six options offered to participants\footnote{The original GZ1 portal has been maintained as a tribute to the great success of the project and can be visited at \url{http://zoo1.galaxyzoo.org/}.}, namely: \textit{Elliptical}; \textit{Clockwise}, \textit{Anti-clockwise} and \textit{Edge-on} Spiral; \textit{Don't Know}; and \textit{Merger}. After the project finalisation, the GZ1 data were published and have since facilitated many follow-up investigations that have resulted in dozens of peer-reviewed publications\footnote{The up-to-date list of Galaxy Zoo publications can be consulted at \url{https://www.zooniverse.org/about/publications}.}.  

There have been numerous attempts to apply ML techniques in astronomy \citep{Ball20101049,Fluke202010}, considering either supervised \citep{Dieleman20151441}, semi-supervised \citep{Rahmani20184416,Slijepcevic2022514}, self-supervised \citep{Hayat2021,Stein2022} or unsupervised \citep{Spindler2020} strategies. Furthermore, it is gaining a great momentum the use of citizen science data in conjunction with unlabelled data in contrastive learning applications \citep{Slijepcevic202202,Walmsley202206}. Within the classification of galaxy images using CNNs, several works have investigated the potential of this approach employing CzS data as well, although neglecting available expert classifications \citep{Dieleman20151441,Kuminski201620,Cheng2021507}. In contrast, the use of CAEs in this particular problem has barely been explored, and have been proposed for feature extraction purposes in supervised classification \citep{Jimenez20208}. However, the joint exploitation of experts, amateurs, and unlabelled data, in conjunction with ML and latest advances in DL has been disregarded across the literature. In the following section, we discuss related SSL approaches and review the basics of DL-based models on which the CzSL methodology is built.

\subsection{Beyond supervised classification with CNNs}
\label{beyond-supervised-classification}
SSL methods aim to extend the advantages of supervised learning by considering large amounts of unlabelled data together with a small number of labelled examples \citep{Zhu20091}. Many approaches have faced this challenge from diverse points of view that can approximately be categorised into generative models \citep{Kingma20143581}, self-training \citep{Triguero201430}, co-training \citep{Appice2017229}, and graph-based SSL approaches \citep{Du20191440}, which have also been coupled with the effectiveness of DL in image classification \citep{Gu2018354}. In our experiments we consider self-training, or more specifically, a self-labelled approach \citep{Triguero2015245}, as a state-of-the-art SSL-based comparative algorithm to test against the proposed CzSL method. These techniques employ unlabelled data within a supervised framework, by which such data are gradually classified, thus enlarging the original training set. We implement a basic version with a CNN that first extends the labels to the entire training data in one single iteration and is then trained again using the full training set, as a comparative approach for our proposed method. 

CNNs \citep{Bengio20091} have been systematically outperforming the benchmarks in image classification over the past few years \citep{Rawat20172352,Gu2018354} due to the increasing development of computational capacities and the availability of large image data sets required for their training \citep{Szegedy20151}. State-of-the-art deep CNNs are often composed of numerous layers \citep{Simonyan2015} and aim to cope with challenging image classification problems \citep{Russakovsky2015211}. In this context, the pre-training and fine-tuning of deep networks \citep{Yosinski20143320}, as a form of transfer learning \citep{Pan20101345}, represents a way of taking advantage of these models in problems where there is a scarcity of labelled examples that completely invalidates their training \citep{Ackermann2018415,Farrens2022}. The augmentation of the training data has also been demonstrated to be effective in improving the learning of CNNs, particularly in astronomy \citep{Dieleman20151441}. By generating more images with slight modifications, such as rotations, flips or small shifts, the network gains generalisation and robustness in predictions and avoids overfitting when training data are scarce \citep{Alhassan20182085,Maslej20211464}. In our model, we leverage a particular implementation of pre-training and fine-tuning that uses amateur and expert labels, respectively, which has already been successfully explored with a simple yet effective CNN \citep{Jimenez20208}. We also investigate the effects of data augmentation applied to our training data sets on the performance of the proposed approach.

\section{CzSL: Learning from experts, citizen science data, and unlabelled data}
\label{CzSL-proposal} 
This section presents the CzSL algorithm. First, we discuss the motivations that led us to devise such an approach to leverage all levels of knowledge referred above: expert- and amateur-labelled data, and unlabelled data (Section \ref{motivation}). Afterwards, we explain the learning methodology (Section \ref{citizen-science-learning}) and describe the selected implementation for the experiments (Section \ref{implementation}).   

\subsection{Motivation: Levels of knowledge}
\label{motivation}
\begin{figure} 
\centering 
\resizebox{0.85\columnwidth}{!}{\includegraphics{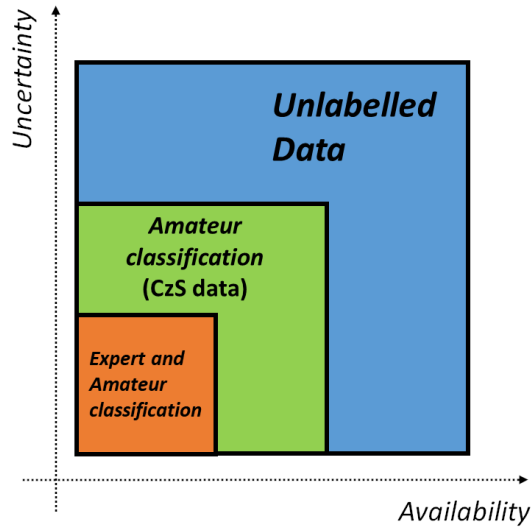}} 
\caption{\label{fig:data-levels-knowledge} Levels of knowledge exploited by the CzSL algorithm.}
\end{figure}

The CzS paradigm offers different levels of knowledge concerning the data being classified that is worth considering in the development of an improved automated classification approach for astronomy, as displayed in Figure \ref{fig:data-levels-knowledge}. On the one hand, we have a pool of amateurs who voluntarily dedicate their time and effort to the execution of an image classification task on the web. They are people with a wide (and unknown) range of capabilities and backgrounds, ranging from children to seniors. Consequently, the resulting amateur-labelled data suffer from an inherent uncertainty that affects its quality, representing the \textit{amateur knowledge} about the training data. On the other hand, a cohort of professional astronomers in the field of study usually validates and complements amateurs' work. They have extensive expertise on the issue, as well as complementary sources of information to support and enrich their judgements. In this paper we will assume that expert ratings provide higher accuracy compared to CzS participants, which means the highest level of knowledge about the data, the so-called \textit{expert knowledge}. In conjunction with expert and amateur classifications, CzSL also delves into the exploitation of unlabelled data, which represent the lowest level of knowledge in this data framework. 

To date, several works have attempted to use CzS data to train automated classifiers, but have typically ignored the potential of both expert-labelled and unlabelled data. This has been developed either in off-line approaches, once the project had been concluded \citep{Banerji2010342,Bahaadini2018172}, or in more recent on-line approaches, whilst the CzS data are being generated \citep{Zevin201734,Walmsley20201554}. Efforts in the former have focused primarily on how to replicate the machine's amateur capability \citep{Banerji2010342,Dieleman20151441}, thus emulating amateur classification skills but also propagating the biases and weaknesses that are latent within CzS data. On the other hand, expert classifications have been leveraged either to validate CzS results \citep{Lintott20081179,Kosmala2016551} or to improve the quality of CzS data \citep{Jimenez2019479}. In the last case, a set of data transformations is able to aggregate valuable information about the uncertainty in amateur classifications by aggregating the information about the uncertainty in amateur classifications held in \textit{Don't Know} scores and the distribution of votes through the classes and examples (\hyperref[appendix-A]{Appendix A}). However, this approach only modifies the amateur scores, which leaves the problem of how to get a final label from multiple rankings unresolved. In addition, the amount of data either labelled by experts or participants in the course of a CzS project are often negligible in comparison with the large volumes of unlabelled data available. 

The proposed CzSL algorithm encompasses the entire set of knowledge levels, establishing a unified framework for learning from both expert and amateur labels, as well as from unlabelled data. To this end, the developed DL-based implementation allows learning from amateur and expert labels by making use of pre-training and fine-tuning of a CNN, respectively, an approach that has been successfully tested in a previous study \citep{Jimenez20208}. However, instead of using amateur and expert classifications separately, CzSL is able to extend the expert knowledge encapsulated in expert classifications to original amateur scores included in CzS data, resulting in a larger amount of high-quality labelled data for the training of the CNN. The complete learning process is described in the following section.    

\subsection{A unified deep learning approach for citizen science}
\label{citizen-science-learning} 
\begin{figure} 
\centering 
\resizebox{1.0\columnwidth}{!}{\includegraphics{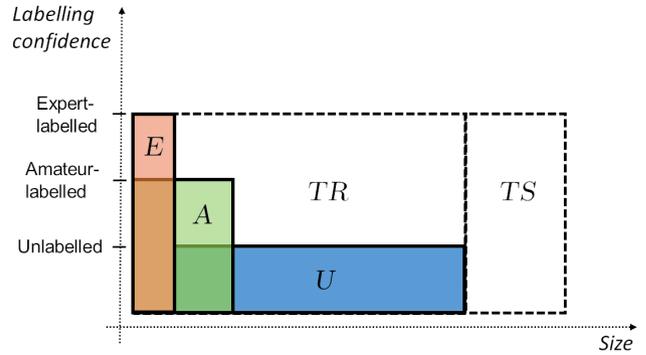}} 
\caption{\label{fig:CzSL-subsets} Data subsets used in the CzSL learning.}
\end{figure}

The data framework employed in the CzSL learning is schematised in Figure \ref{fig:CzSL-subsets}. Let $ TR $ be the training set established to learn from expert classifications, CzS data and unlabelled data, and $ TS $ the test set reserved for testing. Within $ TR $, we consider an \textit{amateur} subset $ A $ of these images as labelled by CzS data, and the rest of the set as unlabelled. That is to say, in $ A $ we keep amateur scores by counting the votes coming from amateur participants, as shown in Table \ref{tab:citizen-science-data}. Additionally, we assume to possess expert classifications for some part of the subset $ A $, an \textit{expert} subset $ E $, thus completing the three levels of knowledge (Figure \ref{fig:data-levels-knowledge}). Note that the $ E $ subset includes images both labelled by experts and amateurs. This is a basic assumption that we make: there are expert classifications available for data already classified in the course of the project. Therefore, $ E \subset A $ by definition. 

CzSL focuses on exploiting the expert-labelled part of the data, which holds the highest confidence, exploring the best way to enlarge the $ E $ set throughout several learning stages that leverage amateur-labelled and unlabelled examples. Hence, in addition to the $ TS $ set, in the subsequent experiments we analyse the prediction capabilities of the model in the portion of expert-unlabelled data in $ TR $, referred to as $ U $. This way, we differentiate between the so-called transductive and inductive learnings, similar to the way it is usually done in SSL experiments. The first one addresses label prediction for the unlabelled part of the set $ TR $ by jointly considering the labelled and unlabelled data provided at the beginning. The latter, however, concerns the whole set $ TR $ as training examples and targets the prediction of unseen data \citep{Triguero2015245}. Therefore, $ E \cup U = TR $, and $ A $ is partially overlapped with $ E $ and $ U $. In terms of the relative size of these sets, it is assumed that $ \vert E \vert \ll \vert A \vert \ll \vert TR \vert $, as it is the case in current CzS projects \citep{Lintott20081179}.        
       
The CzSL methodology consists of three phases that sequentially learn from unlabelled data, amateur-labelled data, and expert-labelled data last. The DL-based implementation proposed enables bottom-up learning: it starts from the bottom up in the knowledge framework described above. This is in contrast to self-labelling strategies in SSL, which extend the labelling from labelled to unlabelled examples \citep{Triguero2015245}. Such approaches tend to make misclassifications in the labelled set enlarging process, which also increases the computational cost in terms of runtime due to repeated trainings. In contrast, CzSL first exploits unlabelled data and then learns from amateur and expert labels. As such, all of the $ TR $ data reinforce the CNN pre-training without using potentially noisy labels across $ U $.  

\begin{figure*}
\centering
\resizebox{1.00\textwidth}{!}{\includegraphics{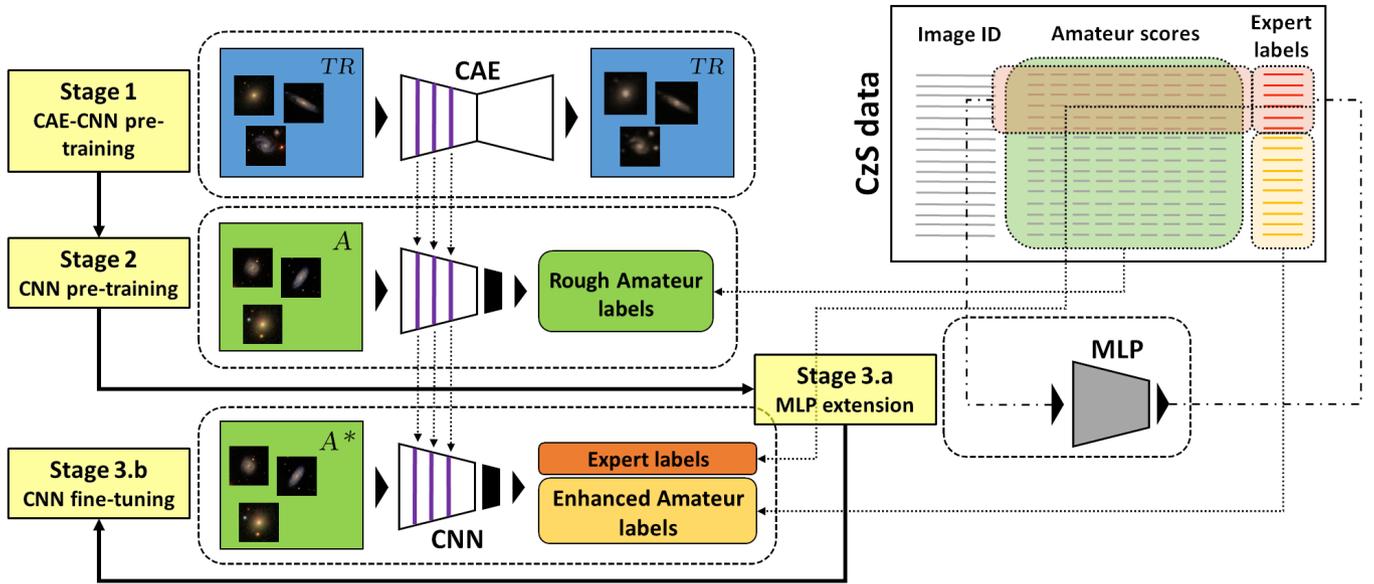}}
\caption{\label{fig:CzSL-workflow} The CzSL learning workflow.}
\end{figure*}

The implementation of CzSL is based on two key elements: the pre-training and fine-tuning of a CNN aided by a CAE, and an MLP. The former is used in the first and second stages of the method to learn from unlabelled and amateur-labelled data using CAE and CNN, respectively, which share a common architecture that makes it possible to transfer the weights across the three stages of the method. The latter is employed in the first part of the third phase to extend the expert knowledge to all of the available CzS data. After this, the CNN is finally fine-tuned, employing expert and enhanced amateur labels. These stages are described below. The complete CzSL learning workflow is depicted in Figure \ref{fig:CzSL-workflow}. 

\begin{itemize}

\item[--] \textbf{Stage 1: CAE-CNN pre-training.} \textit{Exploiting unlabelled data with a CAE}. CzSL starts by employing the whole set of images in $ TR $ with no consideration to image labels. The CAE learns patterns from the data, and as such enables the unsupervised pre-training of the CNN employed in the subsequent stages. As described above, the feature extraction layers of the CNN must match the CAE structure, allowing weights to be transferred between both DL models (dashed arrows from Stage 1 to Stage 2 in Figure \ref{fig:CzSL-workflow}). We refer to this process as CAE-CNN pre-training. 

\item[--] \textbf{Stage 2: CNN pre-training.} \textit{Learning from rough amateur labels}. The CNN, previously loaded layer by layer with the CAE weights, is pre-trained over the $ A $ subset using the amateur labels derived from the scores included in the CzS data. These labels are generated adopting a simple majority criterion: the class holding the highest score is assigned to the example. In case of a tie, the class is assigned at random. This phase aims at pre-training the CNN with \textit{rough} labels, which are more numerous than their expert counterparts although noisier as well, in the sense of potentially incorrect. Then, CNN weights will be transferred to the last stage of the learning process. Here, the CzS data are first leveraged in a rough mode (majority criterion), in contrast with the following phase that focuses on enhancing amateur knowledge with the help of expert classifications.      

\item[--] \textbf{Stage 3.a: MLP extension.} \textit{Extending expert knowledge to CzS data}. At this stage, we complete the extension of the expert knowledge in $ E $ to the remaining examples in $ A $ (note that $ E \subset A $, so this affects the examples in $ A \setminus E $). To this end, we implement an MLP that learns the correlation between expert labels and amateur scores included in the CzS data. The MLP is trained using the expert-labelled portion of the data: amateur scores are used as the MLP input, and expert labels as output. However, note that this stage does not depend on the selected algorithm to carry out the knowledge extension process, and other suitable classifiers could be applied successfully. 

\item[--] \textbf{Stage 3.b: CNN fine-tuning.} \textit{Learning from expert and enhanced amateur labels}. Once the expert knowledge extension has been completed using the amateur scores and expert labels in $ E $, the MLP predictions are employed as enhanced amateur labels for the rest of images in $ A $. Thus, the $ A $ image subset is labelled with expert labels for examples in $ E $, and enhanced amateur labels for examples in $ A \setminus E $, which we denote as $ A^{*} $ from now on. The CNN is loaded layer by layer with the resulting weights of Stage 2 and using identical configuration (dashed arrows from Stage 2 to Stage 3.b in Figure \ref{fig:CzSL-workflow}) is trained on the $ A^{*} $ subset. This completes the CNN fine-tuning with the finest labels, leveraging the highest level of knowledge achieved from expert labels and amateur scores.          

\end{itemize}
After completing the whole process, the CNN has captured the information present in expert labels, rough amateur labels, and patterns learnt in Stage 1 and it is ready to predict unseen examples in $ TS $ set. Although the CzSL integrates both amateur and expert levels of knowledge, we are also interested in exploring how the treatment of the uncertainty by the data transformations proposed in \citet{Jimenez2019479} and briefly introduced in \hyperref[appendix-A]{Appendix A} behave in conjunction with the CzSL learning. To this end, we employ either original or transformed scores as input to the MLP in the Stage 3.a, an aspect that will be tested in the experiments with binary and multi-class classification data sets. In the following section, we detail the implementation of the DL models involved.   

\subsection{Implementation}
\label{implementation}
The CzSL implementation is based on the pre-training and fine-tuning of a CNN using a CAE, and an MLP for expert knowledge extension to CzS data. These architectures are not proposed as a final design for CzSL's implementation, although they have been tested with success in a previous study with GZ1 data that investigated the pre-training and fine-tuning of a CNN employing amateur and expert labels, respectively \citep{Jimenez20208}. On the basis of these results, we add the CAE-CNN pre-training process to exploit the wealth of unlabelled data. However, other CNN or MLP topologies could be beneficial depending on the classification problem being tackled or the data that are used. In what follows, the implementations of the DL models later employed in the experiments are introduced. 

\subsubsection{CAE-CNN Pre-training and CNN Fine-tuning}
We proposed very simple yet effective CAE and CNN models that have been thoroughly tested with success in previous investigations within the galaxy image classification problem \citep{Jimenez20208}. The CAE deploys three pairs of convolution -- pooling layers between input and encoding. The first convolution layer contains 16 kernels, and the second and third hold 8 kernels. Receptive fields are 3$\times$3 pixels size with stride 1 and zero padding, and pooling layers implement max pooling with 2$\times$2 pixels windows and stride 2. ReLU are selected as activation functions along the whole architecture except in the output layer that uses sigmoid function.

The implemented CNN resembles the CAE described above. It is composed of three pairs of convolution -- pooling layers that complete the feature extraction, followed by two dense layers of 256 and 128 neurons and the output layer with same number of neurons as classification classes defined in the problem. For convolution layers, it also computes 16 feature maps in the first layer, and 8 in second and third layers with 3$\times$3 receptive fields in all cases. Pooling layers implement max pooling with 2$\times$2 pixels windows. ReLU activations are used except for the output layer, which uses the SoftMax function, thus giving rounded probability distributions to produce the final class labels. These specifications are summarised in Table \ref{tab:CNN-CAE}. 

\begin{table*} 
\centering
\resizebox{0.8\textwidth}{!}{
\begin{tabular}{cccc||cccc}
\hline
\textbf{Layer Type} & \textbf{Size} & \textbf{Stride} & \textbf{Activation} & \textbf{Layer Type} & \textbf{Size} & \textbf{Stride} & \textbf{Activation} \\\hline
Convolution & 16 kernels (3$\times$3) & 1 & ReLU & Convolution & 16 kernels (3$\times$3) & 1 & ReLU \\ 
Pooling & (2$\times$2) & 2 & -- & Pooling & (2$\times$2) & 2 & -- \\
Convolution & 8 kernels (3$\times$3) & 1 & ReLU & Convolution & 8 kernels (3$\times$3) & 1 & ReLU \\ 
Pooling & (2$\times$2) & 2 & -- & Pooling & (2$\times$2) & 2 & -- \\
Convolution & 8 kernels (3$\times$3) & 1 & ReLU & Convolution & 8 kernels (3$\times$3) & 1 & ReLU \\ 
Pooling & (2$\times$2) & 2 & -- & Pooling & (2$\times$2) & 2 & -- \\
\multicolumn{4}{c||}{\textbf{(Embedded layer})} & Fully connected & 256 neurons & -- & ReLU \\ 
Deconvolution & 8 kernels (3$\times$3) & 1 & ReLU & Fully connected & 128 neurons & -- & ReLU \\  
Pooling & (2$\times$2) & 2 & -- & Output & 2 \textit{or} 3 neurons & -- & SoftMax \\ 
Deconvolution & 8 kernels (3$\times$3) & 1 & ReLU & & & & \\
Pooling & (2$\times$2) & 2 & -- & & & & \\
Deconvolution & 16 kernels (3$\times$3) & 1 & Sigmoid & & & & \\\hline												
\end{tabular}}
\caption{\label{tab:CNN-CAE} CAE (left) and CNN (right) architectures implemented.}
\end{table*} 

\subsubsection{MLP for expert knowledge extension}
MLPs represent the simplest type of feed-forward artificial neural networks \citep{Lecun2015436}, consisting of input and output layers of neurons with a variable number of hidden layers in between. All neurons deployed through the layers are densely connected, and the structure is trained by the back-propagation algorithm \citep{Rumelhart1986533}. The MLP implementation used in our experiments is composed of four hidden layers with 8, 7, 5, and 3 neurons. As with the CNN implemented, SoftMax is applied to the last layer to provide class probabilities. Prior to the experiments, other architectures were thoroughly tested through a grid search that highlighted the model that performed best with the classification problems covered in our experiments. The procedure followed is explained in detail in \hyperref[appendix-B]{Appendix B}.

\section{Experiments}
\label{experiments}
This section presents the experiments that have been conducted to test the proposed CzSL approach. First, we describe the data sets used (Section \ref{datasets}). Then, we introduce a set of comparative approaches defined to develop a well-grounded analysis of the CzSL performance (Section \ref{comparative-approaches}). Finally, we specify the criteria adopted in the evaluation of experiments as well as the implementation parameters employed (Section \ref{experimental-evaluation-parameters}).          

\subsection{Data sets}
\label{datasets}
In the experiments, we leverage three distinct problems derived from the GZ1 project results aiming to test the proposed learning methodology in varied classification environments. The data sets employed are defined by the availability of expert labels, which are considered as ground truth in the evaluation of the results. These data sets are defined as follows, and their features are summarised in Table \ref{tab:datasets}:
\begin{itemize}

\item \textbf{GZ1-(SE)}: This data set derives from the expert catalogues originally employed by the GZ1 developers for the assessment of the project results \citep{Lintott20081179}. These are the MOSES catalogue \citep{Schawinski20071415}, for elliptical galaxies (E), and the Longo catalogue \citep{Longo2011224}, for spirals (S). The combination of both samples forms a total of 40 902 examples classified as spiral or elliptical (binary classification) which is called GZ1-(SE).  

\item \textbf{GZ1-(RLE)}: The Longo catalogue also provides two sub-classifications for spiral galaxies depending on the handedness of its spiral arms. This information enables us to define a multi-class problem by splitting the spiral galaxies included in GZ1-(SE) into two classes: right-handed (R) and left-handed spiral (L). In conjunction with the elliptical class, we define the GZ1-(RLE) data set, also consisting of 40 902 examples.

\item \textbf{GZ1-(MG)}: This data set comes from an expert catalogue generated for the study of galaxy mergers within the original GZ1 data. This is the catalogue published in \citet{Darg20101043}, which includes a set of 3003 visually selected pairs of merging galaxies. We first cross-match this catalogue against the combination of MOSES and Longo catalogues. Galaxies included in both samples are taken as merger\footnote{This is not a contradiction, since a galaxy merger is a fusion between two or more galaxies itself, it might be labelled as either spiral or elliptical with validity depending on the morphology of the most prominent component of the merger.} (MG), whereas the rest of examples are kept as no-merger (No-MG). Then, the rest of examples in the Darg catalogue are added to the final sample as MG. Specifically, there is an overlap of 126 examples both included in GZ1-(SE) and Darg catalogue, and 2877 examples not present in GZ1-(SE) that are added. We refer to this imbalanced data set as GZ1-(MG), which consists of 43 779 examples.  
 
\end{itemize}  
\begin{table}[!h]
\centering
\resizebox{1.0\columnwidth}{!}{
\begin{tabular}{lccc}

\hline 
\textbf{Data set} & \textbf{No. examples} & \multicolumn{2}{c}{\textbf{Classes distribution (\%)}} \\  
                 &                       &     Expert labels & Amateur labels \\\hline
\textbf{GZ1-(SE)} & 40 902 & 60.7 : 39.3 & 58.6 : 41.4 \\ 
\textbf{GZ1-(RLE)} & 40 902 & 30.1 : 30.6 : 39.3 & 28.8 : 29.9 : 41.3 \\ 
\textbf{GZ1-(MG)} & 43 779 & 6.5 : 93.5 & 6.5 : 93.5 \\\hline  
											
\end{tabular}}
\caption{\label{tab:datasets} Data sets used in the experiments. The distribution values follow the same order as in the data set label, e.g. for GZ1-(RLE) the numbers indicate the percentage of right-handed and left-handed spiral, and elliptical classes.}
\end{table} 

The amateur classifications considered are part of the so-called GZ1 \textit{Table 2} data set\footnote{These data are publicly available at \url{https://data.galaxyzoo.org/}.} (GZ1-T2) \citep{Lintott2011166}, which does not provide a final classification for the galaxies but a set of votes for each of the images displayed on the GZ1 website. Previous uses of these data have drawn on the application of different thresholds over these scores in order to obtain samples of varied quality, leaving the major part of the CzS results unused \citep{Jimenez2019479}. In contrast, in this paper we apply a ``strict majority'' criterion to obtain the rough amateur labels required by CzSL, allowing us to exploit the totality of GZ1 results. In the case of a tie between the scores (e.g. elliptical and spiral scores both equal 0.5), we assign a random class to the example. In GZ1-(RLE) and GZ1-(MG) samples, the majority criterion is applied consistently to obtain amateur labels: right-handed spiral and left-handed spiral classes are assigned to the class with the highest score between Clockwise and Anti-clockwise spiral scores, for examples whose combined spiral score is greater than their elliptical score. Likewise, the merger class is allocated to examples holding a merger score greater than 0.5, and ties are handled assigning a random class. 

\begin{figure*}
\centering 
\resizebox{0.80\textwidth}{!}{\includegraphics{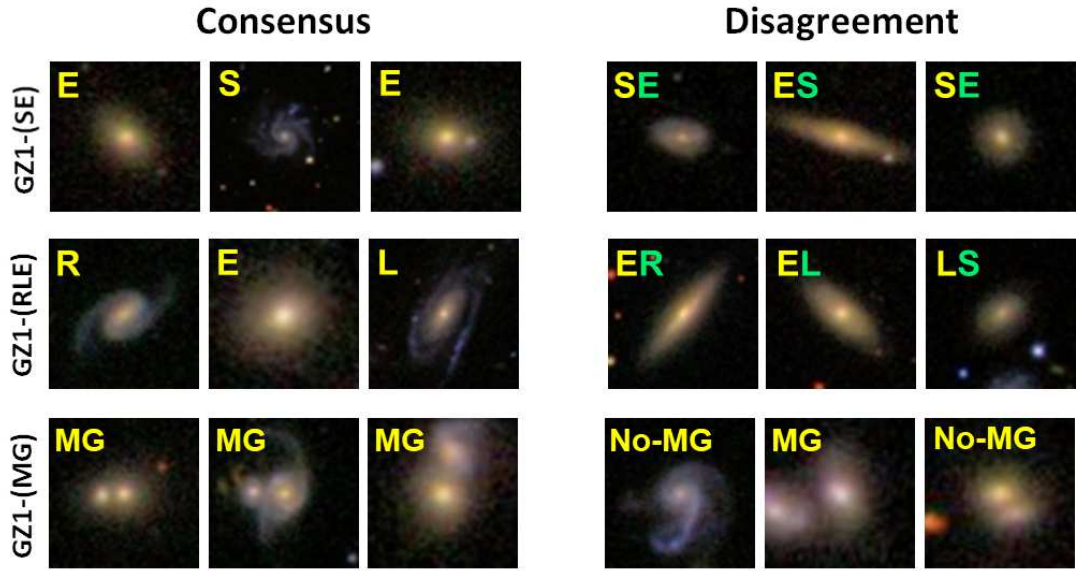}} 
\caption{\label{fig:images-sample} Sample of 64$\times$64 pixels RGB images from the three data sets used in the experiments. Left panel shows images in which there was consensus between expert and amateur classifications. Right panel shows images in which there was disagreement between experts and amateurs: yellow and green labels indicate expert and amateur classifications, respectively.}
\end{figure*}

The images utilised follow the same specifications originally employed in GZ1. We accessed the GZ1-T2 image collection from the Sloan Digital Sky Survey\footnote{\url{https://www.sdss.org/}} using the CAS\footnote{\url{http://cas.sdss.org}} tool. To do so, we followed the original specifications and 424$\times$424 pixels size as is explained in \citet{Lintott20081179}. However, previous investigations concluded that the performance of the proposed CAE and CNN is not significantly affected by the image size \citep{Jimenez20208}. Similarly as is proposed in other works \citep{Dieleman20151441}, we first converted the images to TIFF format and cropped them to half their original size, keeping the galaxy in the centre, and compressed them to 64$\times$64 pixels size to accelerate the training of the DL models. Figure \ref{fig:images-sample} shows a sample of actual images used in the experiments from the three data sets investigated. 

\subsection{Comparative approaches}
\label{comparative-approaches}
Although there is a plethora of approaches that delves into the synergy between unlabelled and labelled data, some of which have been listed above, our proposal focuses on the simultaneous exploitation of the three levels of knowledge identified (Figure \ref{fig:data-levels-knowledge}), which has been barely explored trough the literature. We first conduct a comparative study of the proposed algorithm versus a set of baseline approaches that are trained using different configurations of the data framework defined above (Figure \ref{fig:CzSL-subsets}). In second place, we complete an ablation study in order to elucidate the contribution of the different learning stages of the methodology (Figure \ref{fig:CzSL-workflow}). In all cases the same CNN model is used, which allows us to analyse to what extent it is possible to predict $ TS $ examples by learning from single or combined expert and amateur label sets and introducing the MLP extension into play. We also define an additional approach inspired by self-labelling techniques \citep{Triguero2015245}. These comparative approaches are described below.
\begin{itemize} 

\item \textit{Oracles}. First, we contemplate an ideal situation in which we would hold either expert or amateur classifications for the whole data. We refer to these approaches as \textit{oracles}, as they are meant to serve as upper-bounds for the rest of the experiments. For both \textbf{\textit{Expert Oracle}} and \textbf{\textit{Amateur Oracle}}, the CNN takes the whole $ TR $ set with expert and amateur labels, respectively, although the evaluation in $ TS $ is always carried out using expert labels (ground truth). These two approaches not only establish upper thresholds for the experimental results, but also serve as an assessment of the CzS data, i.e. amateur labels should lead to a worse performance in terms of prediction accuracy. 

\item \textit{Pure supervised learning}. We then train using expert and amateur labels, either independently or jointly. As such, we investigate the performance of the selected CNN in a purely supervised training framework. This will also provide information about the relative difficulty of the classification problems explored and the potential improvements obtained by the CzSL model by treating the results of the pure supervised training as lower-bounds. We define the \textbf{\textit{Expert}}, \textbf{\textit{Amateur}}, and \textbf{\textit{Expert-Amateur}} approaches, in which the CNN learns from $ E $, $ A $, and $ E \cup A $ sets, respectively. In the last case, expert labels are considered for examples in $ E $ and amateur labels for the rest ($ A \setminus E $).

\end{itemize}
We also test the performance of the proposed MLP regardless of the CzSL model in two additional comparative approaches. Recall that CzSL learns patterns from unlabelled data, and then from amateur and expert labels, and MLP predictions, leveraging the full range of available knowledge in a bottom-up fashion. Conversely, these two comparative approaches represent top-down strategies: they learn from expert labels to enhance amateur labels and then extend the labelling to unlabelled data. 
\begin{itemize}

\item \textit{MLP}. The first one, referred to as \textbf{\textit{MLP}}, is intended to investigate the MLP capabilities to replicate expert labels by using amateur scores only, enlarging the $ E $ set with additional examples present in $ A $. The MLP is trained on $ E $ using expert labels, and then employed to enhance amateur labels in $ A \setminus E $. After this, the CNN is trained on $ A^{*} $ using expert labels and MLP predictions for images in $ E $ and $ A \setminus E $, respectively. 

\item \textit{MLP Self-labelling}. The second approach is a test of the MLP within a different learning algorithm, allowing us to evaluate this component of the CzSL model in a different setting. This approach, referred to as MLP-Self-labelling and designated from now on as \textbf{\textit{MLP-SL}}, extends the previous MLP algorithm explained above to a basic self-labelled approach after completing the extension of expert knowledge. The CNN is trained over the entire $ TR $ set using expert labels ($ E $), MLP predictions ($ A \setminus E $), and the CNN predictions for the rest of the examples in $ U $, that is, 75\% of the data are not included either in $ E $ nor $ A $. For simplicity, we only complete one single iteration, given that multiple iterations would invalidate such an approach in terms of the training time. 

\end{itemize}
The complete set of comparative approaches is summarised in Table \ref{tab:comparative-approaches}. With MLP and MLP-SL approaches we also investigate the use of the data transformations developed in \citet{Jimenez2019479}, which aggregate additional information about the uncertainty present in the original GZ1-T2 data set. We compare the performance of these approaches employing either the transformed scores or the original ones in GZ1-(SE) and GZ1-(RLE) data sets. The scores inputted to the MLP in Stage 3.a are those for the \textit{Elliptical}, \textit{Combined Spiral}, \textit{Clockwise Spiral}, \textit{Anti-clockwise Spiral}, \textit{Edge-on Spiral}, \textit{Don't Know}, and \textit{Merger} classes. Therefore, with the transformed scores, the elliptical and combined spiral original scores were replaced by the scores obtained after applying the data transformation that best performed in \citet{Jimenez2019479}. This transformation sequence is composed of a combination of \textit{Normalisation}, \textit{DK Votes shift}, and \textit{Votes boost} transformations, applied to the scores in this order. A detailed explanation is included in \hyperref[appendix-A]{Appendix A}.

\begin{table*}[!h]
\centering
\resizebox{1.0\textwidth}{!}{
\begin{tabular}{llll}

\hline       
\multicolumn{1}{c}{\textbf{Approach}} & \textbf{Learns from} (\% of $ TR $) & \textbf{Labels used} & \multicolumn{1}{c}{\textbf{Description}} \\\hline  
 
\textbf{\textit{Expert Oracle}} & $ TR $ (100\%) & Expert & CNN trained over the entire $ TR $ set with expert labels \\ 

\textbf{\textit{Amateur Oracle}} & $ TR $ (100\%) & Amateur & CNN trained over the entire $ TR $ set with amateur labels \\

\textbf{\textit{Expert}} & $ E $ (5\%) & Expert & CNN trained over the $ E $ set with expert labels \\ 

\textbf{\textit{Amateur}} & $ A $ (25\%) & Amateur & CNN trained over the $ A $ set with amateur labels \\ 

\textbf{\textit{Expert-Amateur}} & $ E \cup A $ (25\%) & Expert and Amateur & CNN trained over $ E \cup A $ using expert labels for examples \\

& & & in $ E $ and amateur labels in the rest ($ A \setminus E $) \\

\textbf{\textit{MLP}} & $ A^{*} $ (25\%) & Expert and MLP predictions & CNN trained over $ A^{*} $ using expert labels for examples \\

& & & in $ E $ and MLP predictions in the rest ($ A \setminus E $) \\

\textbf{\textit{MLP-SL}} & $ TR $ (100\%) & Expert, MLP predictions, & CNN first trained over $ A^{*} $ using expert labels ($ E $) and \\

& & and CNN predictions & MLP predictions ($ A \setminus E $), and then over the entire $ TR $  \\
 
& & & adding its own predictions in the rest of examples ($ U \setminus A $) \\\hline 
								   	
\end{tabular}}
\caption{\label{tab:comparative-approaches} Comparative approaches established for the evaluation of the CzSL model.}
\end{table*}   

For the experiments with the GZ1-(MG) data set, we consider two additional settings in order to verify the experimental results without the added complication of imbalanced classification. Amongst the plethora of data pre-processing techniques that have been widely studied in the literature \citep{Thabtah2020429}, we opt for random over-sampling (ROS) and random under-sampling (RUS), two simple approaches that have extensively provided good results \citep{Galar2012463}. ROS and RUS are sampling techniques that most directly adjust the imbalance ratio (IR), defined as the ratio between the majority and minority classes numbers of examples in the data \citep{Lopez2013113}. For ROS, instances of the minority class are randomly replicated until certain IR is achieved. In the case of RUS, examples of the majority class are randomly removed to balance the IR. In our experiments ROS and RUS are applied to reach IR = 1.0, that is, the same number of merger and no-merger examples in the final data.   

\subsection{Experimental evaluation and parameters}
\label{experimental-evaluation-parameters} 
The data sets were partitioned using a 5-fold cross-validation scheme, so that 80\% of the data were employed as a $ TR $ set and the remaining 20\% as a $ TS $ set. In a similar way to how SSL experiments are usually conducted \citep{Triguero2015245}, $ TR $ is composed of a small amount of labelled data and a larger set of unlabelled data. Following the nomenclature previously introduced (Section \ref{citizen-science-learning}), 75\% of the examples are included in the $ U $ set and taken as unlabelled (labels are ignored), and the rest (25\%) is assumed to hold amateur labels based on CzS data, forming the $ A $ set. Of the data included in $ A $, we consider 5\% of the entire $ TR $ as expert-labelled ($ E $). Therefore, CzSL takes the $ E $ set entirely included in $ A $: we assume that we have expert classifications for 20\% of the data labelled by amateurs. These percentages are arbitrary and are only meant to reflect the relative abundancy of expert-, amateur-labelled, and unlabelled images. Mathematically: $ E \cup U = TR $, and $ E \subset A $. The sizes of these sets remain fixed within each of the five $ TR $ partitions arranged as a result of the 5-fold cross-validation, allowing us to keep the number of experiments to a reasonable amount. Nevertheless, future work will entail the exploration of 25\% and 5\% random samples of $ TR $ to produce the $ A $ and $ E $ sets, respectively. We first sample the 5-fold cross-validation and then define $ E $, $ A $, and $ U $ sets from $ TR $.           
 
Expert classifications are always regarded as ground truth for the evaluation of results, although classifications carried out by expert astronomers are also prone to be biased or incorrect. However, we will assume that expert classifications are the most reliable in terms of classification accuracy. For GZ1-(SE) and GZ1-(RLE) data sets, given that the class distributions are well-balanced (Table \ref{tab:datasets}), we employ the accuracy (Acc) metric to compare the performance of the different approaches \citep{Witten2005}. Acc reports the number of correctly predicted examples over the total number of classified examples. However, this measure is less representative in imbalanced classification frameworks, given that the classes are not equally weighted in such cases. Instead, we use the geometric mean (G-mean) \citep{Lopez2013113} for the experiments involving the GZ1-(MG) data set. G-mean is calculated as the geometric mean of the true positive rate ($ TP_{rate} $) and true negative rate ($ TN_{rate} $), which reports how well the algorithm examined is able to predict both classes at the same time. $ TP_{rate} $ and $ TN_{rate} $ quantify the percentage of positive and negative instances correctly classified, respectively, and are mathematically defined as follows: 

\[ TP_{rate} = \dfrac{TP}{TP + FN} , \;\; TN_{rate} = \dfrac{TN}{FP + TN} \, .\] 

Then: 

\[ \text{G-mean} = \sqrt{TP_{rate} \cdot TN_{rate}} \, .\]    

Besides the classification accuracy, we also examine the runtime to characterise the performance of each of the comparative approaches defined. For Expert, Amateur, and Expert-Amateur approaches, the runtime is expected to be proportional to the amount of data used in the model's training. However, for MLP, MLP-SL, and CzSL this comparison is not so straightforward, given that these models are composed of several phases employing different partitions of the $ TR $ set. In order to minimise the effect of the randomness due to the initialisation of weights in DL models, we completed ten executions of every single experiment. Hence, the metrics' values presented in what follows correspond to the average over ten independent executions of the same experiment. We take as representative the average runtime of the training over one single partition of the 5-fold cross-validation ($ TR $ data) along with its standard deviation. In this respect, we report the CNN training time only, as this is the more significant phase of the experiment in terms of total runtime and accounts for a stage that is common to every approach across the three data sets investigated. 

All the experiments were run in a single node with an Intel(R) Xeon(R) CPU E5-1650 v4 processor (12 cores) at 3.60GHz and 64 GB of RAM. The training of CAE and CNN models implemented either in the CzSL approach or other algorithms used a NVIDIA Titan Xp GPU. CAE and CNN were trained over 100 epochs with a batch size of 256 images, using mini-batch gradient descent for optimisation, and $ TR $ was split into 70/30 for training and validation, respectively. The MLP was trained over 300 iterations, employing the ReLU activation function and stochastic gradient descent as well. In terms of software, we used the \textsc{keras}\footnote{\url{https://keras.io/}} and \textsc{scikit-learn}\footnote{\url{https://scikit-learn.org}} Python packages for the implementation of CAE and CNN models, and evaluation metrics. Except for the aforementioned specifications, the remaining parameters are kept at their default values.

\section{Results and analysis}
\label{results-analysis}
This section presents the results of the experiments conducted for the testing of the proposed CzSL approach in accordance with the comparative approaches and experimental framework that are described in the previous section. The obtained results along with their analysis are summarised in the following for oracles and the three data sets covered (Sections \ref{oracles}, \ref{binary-multiclass}, and \ref{imbalanced}, respectively). After this, we investigate the application of data augmentation techniques on the training data sets, comparing their effects on Expert and Amateur comparative approaches, as well as on CzSL (Section \ref{data-augmentation}). Finally, the CzSL's learning process is examined in an ablation study that elucidates the contribution of each of the learning phases of the algorithm to its overall performance (Section \ref{ablation-study}).

\subsection{Oracles}
\label{oracles}
\begin{figure}
\centering 
\resizebox{1.0\columnwidth}{!}{\includegraphics{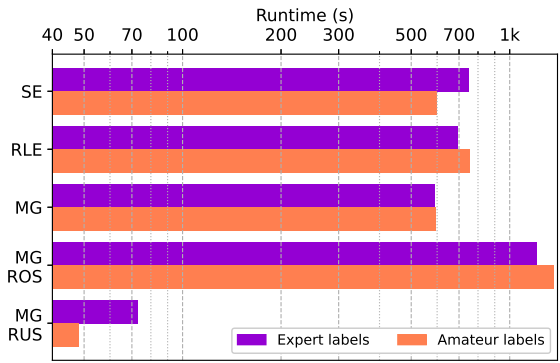}} 
\caption{\label{fig:oracles-runtimes} Execution times in logarithmic scale for oracles.}
\end{figure}

Oracles employ the entire $ TR $ set for completing the CNN training in a pure supervised way, either using expert or amateur labels. These results are shown in Tables \ref{tab:oracles-results} for Acc and G-mean average values obtained in $ TS $ set, and Figure \ref{fig:oracles-runtimes} for average training times for the CNN over one single $ TR $ partition (average values over the ten executions completed in all cases). The computational time taken by ROS and RUS pre-processing phases is considered negligible. 

\begin{table}[!h]
\centering
\resizebox{0.85\columnwidth}{!}{
\begin{tabular}{lcc}

\hline
\textbf{Data set} & \textbf{Expert Oracle} & \textbf{Amateur Oracle} \\
                  & Acc\textsubscript{\textit{TS}} & Acc\textsubscript{\textit{TS}}  \\\hline
        
\textbf{GZ1-(SE)}  & 0.9510 $\pm$ 0.0037 & 0.9169 $\pm$ 0.0053 \\ 
\textbf{GZ1-(RLE)} & 0.8181 $\pm$ 0.0166 & 0.8065 $\pm$ 0.0165 \\ 

& G-mean\textsubscript{\textit{TS}} & G-mean\textsubscript{\textit{TS}} \\\hline
\textbf{GZ1-(MG)}  & 0.7592 $\pm$ 0.0274 & 0.5768 $\pm$ 0.0613 \\        
\textbf{GZ1-(MG) + ROS} & 0.8487 $\pm$ 0.0026 & 0.7669 $\pm$ 0.0048 \\  
\textbf{GZ1-(MG) + RUS} & 0.8738 $\pm$ 0.0092 & 0.8754 $\pm$ 0.0050 \\\hline 
								   	
\end{tabular}}
\caption{\label{tab:oracles-results} Results for oracles in the three data sets investigated.}
\end{table} 

Although the three classification problems employ the same sort of images, experimental results show a wide range of difficulty in the recognition of patterns by the CNN. The CNN shows the best performance with the binary problem. Then distinction between right-hand and left-hand spirals appears to be more difficult, and the identification of galaxy mergers highlights as the hardest problem. There is also a big difference in terms of label quality that is clearer in GZ1-(SE) and GZ1-(MG) data sets, probably because the identification of handedness does not depend on experienced knowledge (less discrepancy between expert and amateur labels). The training using expert labels yields better results, except for classification of mergers with RUS. This may be due to the inherent randomness of the under-sampling process in conjunction with the small amount of data employed, which could tend to diminish the distinction between amateur and expert labels when the number of examples used in the learning is minimal. The pre-processing techniques used with the GZ1-(MG) data set also have a big impact in terms of G-mean and runtime: RUS provides the best performance employing the shortest runtime.           

\subsection{Binary and multi-class classification}
\label{binary-multiclass}

\begin{table*}[!h]
\centering
\resizebox{0.75\textwidth}{!}{
\begin{tabular}{lcc||cc}

\hline 
\multicolumn{1}{c}{\textbf{Approach}} & \multicolumn{2}{c}{\textbf{GZ1-(SE)}} & \multicolumn{2}{c}{\textbf{GZ1-(RLE)}} \\  
                                       & Acc\textsubscript{\textit{U}} & Acc\textsubscript{\textit{TS}} & Acc\textsubscript{\textit{U}} & Acc\textsubscript{\textit{TS}} \\\hline

\textbf{\textit{Expert Oracle}} & -- & 0.9510 $\pm$ 0.0037 & -- & 0.8181 $\pm$ 0.0166 \\ 
\textbf{\textit{Amateur Oracle}} & -- & 0.9169 $\pm$ 0.0053 & -- & 0.8065 $\pm$ 0.0165 \\\hline
                                    
\textbf{\textit{Expert}} & 0.9222 $\pm$ 0.0040 & 0.9225 $\pm$ 0.0047 & 0.6320 $\pm$ 0.0058 & 0.6367 $\pm$ 0.0071 \\ 
\textbf{\textit{Amateur}} & 0.9162 $\pm$ 0.0028 & 0.9143 $\pm$ 0.0044 & 0.6520 $\pm$ 0.0074 & 0.6468 $\pm$ 0.0055 \\
\textbf{\textit{Expert-Amateur}} & 0.9288 $\pm$ 0.0033 & 0.9285 $\pm$ 0.0038 & 0.6595 $\pm$ 0.0037 & 0.6555 $\pm$ 0.0033 \\
\textbf{\textit{MLP Original}} & 0.9392 $\pm$ 0.0026 & 0.9381 $\pm$ 0.0023 & 0.6601 $\pm$ 0.0022 & 0.6563 $\pm$ 0.0030 \\
\textbf{\textit{MLP Transformed}} & 0.9411 $\pm$ 0.0017 & 0.9402 $\pm$ 0.0023 & 0.6622 $\pm$ 0.0054 & 0.6585 $\pm$ 0.0042 \\
\textbf{\textit{MLP-SL Original}} & 0.9419 $\pm$ 0.0016 & 0.9407 $\pm$ 0.0018 & 0.6735 $\pm$ 0.0059 & 0.6665 $\pm$ 0.0057 \\
\textbf{\textit{MLP-SL Transformed}} & \textbf{0.9440 $\pm$ 0.0018} & \textbf{0.9430 $\pm$ 0.0023} & 0.6733 $\pm$ 0.0065 & 0.6656 $\pm$ 0.0058 \\
\textbf{\textit{CzSL Original}} & 0.9396 $\pm$ 0.0025 & 0.9372 $\pm$ 0.0029 & 0.7334 $\pm$ 0.0062 & 0.7167 $\pm$ 0.0061 \\
\textbf{\textit{CzSL Transformed}} & 0.9435 $\pm$ 0.0028 & 0.9405 $\pm$ 0.0036 & \textbf{0.7380 $\pm$ 0.0049} & \textbf{0.7192 $\pm$ 0.0052} \\\hline
								   	
\end{tabular}}
\caption{\label{tab:ES-ERL-results} Results with GZ1-(SE) and GZ1-(RLE) data sets. In each column, the best result is highlighted.}
\end{table*}

\begin{table*}[!h]
\centering
\resizebox{1.0\textwidth}{!}{
\begin{tabular}{ll||*3c||*4c}

\hline         

\multicolumn{1}{c}{\textbf{Data set}} & \multicolumn{1}{c}{\textbf{MLP input scores}} & \multicolumn{4}{c}{\textbf{MLP-SL}} & \multicolumn{3}{c}{\textbf{CzSL}} \\

& & Training ($ E \cup A^{*} $) & Training ($ TR $) & Total & CAE Training ($ TR $) & Pre-training ($ E \cup A $) & Fine-tuning ($ E \cup A^{*} $) & Total \\\hline


\multirow{2}{*}{\textbf{GZ1-(SE)}} & Original & 272.0 $\pm$ 3.9 & 1185.0 $\pm$ 16.5 & 1457.0 $\pm$ 20.4 & \multirow{4}{*}{1422.3 $\pm$ 65.4} & 207.7 $\pm$ 7.7 & 209.0 $\pm$ 7.5 & 1839.0 $\pm$ 80.6 \\ 

& Transformed & 264.5 $\pm$ 4.1 & 1180.5 $\pm$ 18.9 & 1445.0 $\pm$ 23.0 & & 236.1 $\pm$ 3.9 & 236.2 $\pm$ 4.7 & 1894.6 $\pm$ 74.0 \\

\multirow{2}{*}{\textbf{GZ1-(RLE)}} & Original & 268.6 $\pm$ 7.1 & 1176.8 $\pm$ 16.3 & 1445.4 $\pm$ 23.4 & & 208.7 $\pm$ 5.3 & 211.5 $\pm$ 4.7 & 1842.5 $\pm$ 75.4 \\

& Transformed & 251.0 $\pm$ 8.4 & 1121.5 $\pm$ 15.7 & 1372.5 $\pm$ 24.1 & & 167.3 $\pm$ 9.2 & 165.0 $\pm$ 6.8 & 1754.6 $\pm$ 81.4 \\\hline
                                   	
\end{tabular}}
\caption{\label{tab:CzSL-MLP-Self-runtimes-ES-ERL} MLP-SL and CzSL total training time analysis for GZ1-(SE) and GZ1-(RLE) data sets. All values are in seconds (s).}

\end{table*} 

\begin{figure}
\centering 
\resizebox{1.0\columnwidth}{!}{\includegraphics{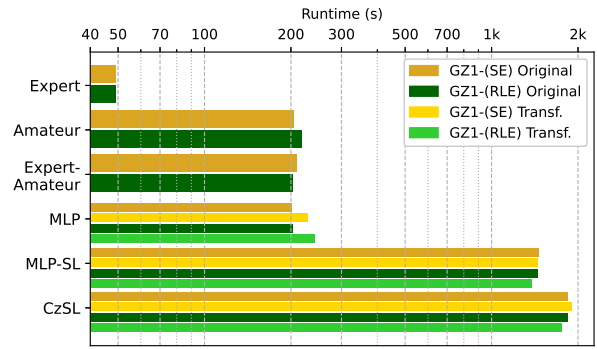}} 
\caption{\label{fig:ES-ERL-runtimes} Execution times in logarithmic scale for GZ1-(SE) and GZ1-(RLE) data sets.}
\end{figure}

The experiments with GZ1-(SE) and GZ1-(RLE) data sets address binary and multi-class classification problems, respectively. In the former case, the CNN is trained to distinguish between elliptical and spiral classes. In the latter, it must also discern the handedness of the (spiral) galaxy's arms. In these two problems, we additionally investigate the influence of the data transformations proposed in \citet{Jimenez2019479}, comparing the original scores, as they are included in GZ1-T2 data, against their transformed version. This comparison only involves the approaches that use the MLP classifier: MLP, CzSL, and MLP-SL. Results are presented in Table \ref{tab:ES-ERL-results} for average Acc values in transductive and inductive learning, that is, the performance in $ U $ and $ TS $ sets, respectively. Figure \ref{fig:ES-ERL-runtimes} depicts the total training times, where top and bottom bars represent the same experiment using original and transformed scores, respectively. In addition, the total training times of MLP-SL and CzSL are detailed in Table \ref{tab:CzSL-MLP-Self-runtimes-ES-ERL}, where the different phases' average runtime is presented.

From these results, we highlight the following remarks:
\begin{itemize}

\item[--] The comparison against oracles shows that there is a wider margin of improvement in the multi-class problem. The improvement provided by the use of more labelled data is almost homogeneous, except when using amateur data only (Amateur approach), starting in the pure supervised approaches, then MLP-based, and CzSL. Once more, the employment of expert labels in comparison with their amateur counterparts seems more important for GZ1-(SE) than for GZ1-(RLE), probably indicating that the noise margin is quite similar for both label sets in the latter problem. 

\item[--] The use of additional levels of knowledge clearly represents an advantage, as the ablation study later confirms. For both data sets, the employment of expert \textit{plus} amateur-based data outstrip these sets separately. The addition of unlabelled data also proves to be key, either by MLP-based approaches or CzSL. In GZ1-(SE), MLP-SL slightly outperforms CzSL, and this result is more accentuated in $ TS $ data. Nonetheless, this improvement is negligible in comparison with the quality jump shown by CzSL in the multi-class setting, which outperforms all top-down strategies explored.
       
\item[--] Transformed scores surpass original ones for both GZ1-(SE) and GZ1-(RLE) data sets, except for the MLP-SL approach for the latter. This can be explained considering that the transformations were designed to be used in the context of binary classification. Thus, the distinction between right-handed and left-handed spiral classes might not be improved by using them. However, the overall tendency points towards the utility of this approach within the CzSL learning, depending on the classification problem covered.     

\item[--] CzSL increases the required runtime with respect to MLP-SL in $ \sim $500 s, due to the CAE training over the entire $ TR $ set (Table \ref{tab:CzSL-MLP-Self-runtimes-ES-ERL}), which signifies a slower training for the proposed CzSL with the data partitions established for the experiments. For the rest of approaches, execution times are similar for both classification problems and proportional to the amount of data used in the training. 

\end{itemize}        
\subsection{Imbalanced classification}
\label{imbalanced}
\begin{figure}
\centering 
\resizebox{1.0\columnwidth}{!}{\includegraphics{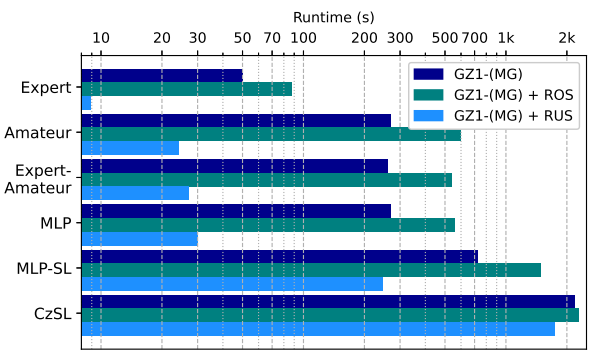}} 
\caption{Execution times in logarithmic scale for GZ1-(MG) data set with no pre-processing, ROS, and RUS techniques.}
\label{fig:MG-runtimes}
\end{figure}

Experiments with the GZ1-(MG) data set tackle the added difficulty of imbalanced classification, for which we provide two additional settings with ROS and RUS techniques. In these experiments, the CNN is trained to distinguish between galaxy mergers and single galaxies, either elliptical or spiral, using original scores only. Results are presented in Table \ref{tab:MG-all} for average G-mean values in $ U $ and $ TS $ sets. Figure \ref{fig:MG-runtimes} depicts the total training times. As above, the total training times of MLP-SL and CzSL are analysed in Table \ref{tab:CzSL-MLP-Self-runtimes-MG}. All values indicated are in seconds.  

\begin{table*}[!h]
\centering
\resizebox{1.00\textwidth}{!}{
\begin{tabular}{lcc||cc||cc}

\hline         
\multicolumn{1}{c}{\textbf{Approach}} & \multicolumn{2}{c}{\textbf{GZ1-(MG)}} & \multicolumn{2}{c}{\textbf{GZ1-(MG) + ROS}} & \multicolumn{2}{c}{\textbf{GZ1-(MG) + RUS}} \\  
& G-mean\textsubscript{\textit{U}} & G-mean\textsubscript{\textit{TS}} & G-mean\textsubscript{\textit{U}} & G-mean\textsubscript{\textit{TS}} & G-mean\textsubscript{\textit{U}} & G-mean\textsubscript{\textit{TS}} \\\hline

\textbf{\textit{Expert Oracle}}  & -- & 0.7592 $\pm$ 0.0274 & -- & 0.8487 $\pm$ 0.0026 & -- & 0.8738 $\pm$ 0.0092 \\ 
\textbf{\textit{Amateur Oracle}} & -- & 0.5768 $\pm$ 0.0613 & -- & 0.7669 $\pm$ 0.0048 & -- & 0.8754 $\pm$ 0.0050 \\\hline
                                    
\textbf{\textit{Expert}}  & 0.2103 $\pm$ 0.0947 & 0.2096 $\pm$ 0.0973 & 0.7802 $\pm$ 0.0087 & 0.7834 $\pm$ 0.0091 & 0.6537 $\pm$ 0.0735 & 0.6537 $\pm$ 0.0736 \\ 
\textbf{\textit{Amateur}} & 0.2650 $\pm$ 0.1061 & 0.2624 $\pm$ 0.1063 & 0.7713 $\pm$ 0.0045 & 0.7482 $\pm$ 0.0066 & 0.8137 $\pm$ 0.0253 & 0.8107 $\pm$ 0.0226 \\
\textbf{\textit{Expert-Amateur}} & 0.4252 $\pm$ 0.0869 & 0.4188 $\pm$ 0.0850 & 0.7951 $\pm$ 0.0130 & 0.7777 $\pm$ 0.0136 & 0.8182 $\pm$ 0.0185 & 0.8158 $\pm$ 0.0194 \\
\textbf{\textit{MLP}} & 0.7054 $\pm$ 0.0391 & 0.6957 $\pm$ 0.0377 & 0.8653 $\pm$ 0.0040 & 0.8333 $\pm$ 0.0069 & 0.8479 $\pm$ 0.0143 & 0.8439 $\pm$ 0.0141 \\
\textbf{\textit{MLP-SL}} & 0.7255 $\pm$ 0.0402 & 0.7053 $\pm$ 0.0422 & \textbf{0.8701 $\pm$ 0.0037} & \textbf{0.8373 $\pm$ 0.0056} & 0.8629 $\pm$ 0.0073 & 0.8573 $\pm$ 0.0086 \\
\textbf{\textit{CzSL}} & \textbf{0.7649 $\pm$ 0.0188} & \textbf{0.7431 $\pm$ 0.0217} & 0.8508 $\pm$ 0.0039 & 0.8144 $\pm$ 0.0049 & \textbf{0.8693 $\pm$ 0.0071} & \textbf{0.8631 $\pm$ 0.0079} \\\hline
								   	
\end{tabular}}
\caption{\label{tab:MG-all} Results with GZ1-(MG) data set applying no pre-processing, ROS, and RUS. In each column, the best result is highlighted.}
\end{table*}

\begin{table*}[!h]
\centering
\resizebox{1.00\textwidth}{!}{
\begin{tabular}{l||*3c||*4c}

\hline         

\multicolumn{1}{c}{\textbf{Data set}} & \multicolumn{4}{c}{\textbf{MLP-SL}} & \multicolumn{3}{c}{\textbf{CzSL}} \\

& Training ($ E \cup A^{*} $) & Training ($ TR $) & Total & CAE Training ($ TR $) & Pre-training ($ E \cup A $) & Fine-tuning ($ E \cup A^{*} $) & Total \\\hline


\textbf{GZ1-(MG)} & 143.5 $\pm$ 3.7 & 584.2 $\pm$ 6.4 & 727.7 $\pm$ 10.1 & \multirow{3}{*}{1712.0 $\pm$ 124.8} & 230.6 $\pm$ 4.7 & 231.1 $\pm$ 6.5 & 2173.7 $\pm$ 136.0 \\ 

\textbf{GZ1-(MG) + ROS} & 290.3 $\pm$ 18.6 & 1191.0 $\pm$ 44.9 & 1481.3 $\pm$ 63.5 & & 291.9 $\pm$ 11.2 & 280.0 $\pm$ 7.7 & 2283.9 $\pm$ 143.7 \\

\textbf{GZ1-(MG) + RUS} & 24.7 $\pm$ 2.0 & 221.0 $\pm$ 43.6 & 245.7 $\pm$ 45.6 & & 13.8 $\pm$ 0.6 & 21.3 $\pm$ 0.2 & 1747.1 $\pm$ 125.6 \\\hline
                                   	
\end{tabular}}
\caption{\label{tab:CzSL-MLP-Self-runtimes-MG} MLP-SL and CzSL total training time analysis for GZ1-(MG) data set applying no pre-processing, ROS, and RUS techniques. All values are indicated in seconds (s).}

\end{table*} 

From these results, we highlight the following remarks:
\begin{itemize}

\item[--] The inclusion of ROS and RUS techniques has proven to be crucial, considering their overall performance in oracles, comparative approaches, and CzSL. As in GZ1-(SE), the best results across the three variants of the problem reach comparable performance to oracles. Employing 5\% and 25\% of $ TR $ data only (with expert and amateur labels, respectively), CzSL and MLP-SL are able to provide similar performance to the CNN using 100\% of the data. This clearly confirms the advantage behind the use of more data and levels of knowledge.     

\item[--] CzSL provides the best result in experiments with no pre-processing and RUS, showing more robustness in the former with respect to the other two MLP-based approaches. However, MLP-SL beats CzSL in the classification with ROS, achieving an accuracy in $ TS $ similar to MLP approach. This might indicate that the CAE-CNN and CNN pre-training phases of CzSL learning lose effectiveness with the over-sampling of the data, by which the repetition of examples might prevent these models from learning more general representations. 

\item[--] Similarly to classification with GZ1-(SE) and GZ1-(RLE), CzSL increases the total runtime in comparison to MLP-SL on account of the CAE training phase across the whole $ TR $ data. Execution times are always proportional to the amount of data employed in the training as a result of the application of ROS and RUS techniques, or the unmodified GZ1-(MG) data set. 

\end{itemize}
\subsection{Data augmentation with CzSL}
\label{data-augmentation}
Previous experiments were carried out considering the data sets with no modifications (Table \ref{tab:datasets}), except for the ROS and RUS pre-processings with GZ1-(MG) images. However, it has been demonstrated in the literature that data augmentation is capable of enhancing the learning of the network when its complexity and the number of trainable weights exceed the available amount of images \citep{Alhassan20182085,Maslej20211464}. To ensure that the results reported above for the Expert and Amateur baselines were sufficiently competitive w.r.t. using other strategies to combat lack of labelled data, we aim to investigate the effect of data augmentation on the CNN's learning that is implemented within CzSL.

We make use of the \texttt{ImageDataGenerator}\footnote{\url{https://keras.io/api/preprocessing/image/}} utility from the \textsc{Keras} Python package, which generates augmented images from a set of training data in a flexible and efficient way. It performs slight variations to the images, like rotations, vertical or horizontal flips, and width or height shifts. The variation ranges are specified for each of the transformations, and new images are generated randomly in a certain amount. This randomness introduces diversity and variation in the training images, which is expected to enhance the CNN's prediction ability \citep{Dieleman20151441}. To select the configuration, we first compute several tests with Expert and Amateur comparative approaches to assess the influence of different transformations and ranges, noticing slight variations in the results. We, therefore, establish a comprehensive parameter setup that includes rotation, width and height shifts, shear, zoom, and horizontal and vertical flips, as follows: \\

\texttt{ImageDataGenerator(rotation\_range=270,} 
\begin{addmargin}[11em]{4em}
\texttt{width\_shift\_range=0.2,} \\
\texttt{height\_shift\_range=0.2,} \\
\texttt{shear\_range=0.2,} \\
\texttt{zoom\_range=0.2,} \\
\texttt{horizontal\_flip=True,} \\
\texttt{vertical\_flip=True,} \\
\texttt{fill\_mode='nearest')} \\
\end{addmargin}

\noindent
By this, augmented images are rotated in a range of 270 degrees, are zoomed up to 20\%, and shifts and shears affect to the 20\% of image extent. Horizontal and vertical flips are also applied. However, in the experiments involving the GZ1-(RLE) data set, these are disabled, because a single horizontal or vertical flip would result in a change in the handedness of the galaxy, essentially converting right-handed spirals into left-handed ones and vice versa. The \texttt{'nearest'} fill mode indicates that new pixels that may be created or modified during transformations are filled with the value of the nearest existing pixel. 

In these experiments, we augment the CNN's training data, maintaining the test and validation partitions unmodified. Using this configuration, we extended the training data to four times its size as it is suggested in related works with similar images \citep{Maslej20211464}. In order to get an initial reference of the effect of the chosen technique, we first implement data augmentation for Expert and Amateur comparative approaches (Table \ref{tab:comparative-approaches}), which serves of a more competitive baseline using only expert and amateur labels, respectively. After this, we check whether the proposed CzSL can also benefit from using data augmentation. We do this in two different ways: augmenting the training data in Stage 2 only and, secondly, in both Stages 2 and 3 (Figure \ref{fig:CzSL-workflow}). These results are shown in Table \ref{tab:data-augmentation} for the three data sets investigated. For simplicity, we only compare performance metrics on $ TS $ data. In experiments with GZ1-(SE) and GZ1-(RLE) we employ the transformed scores, which in previous experiments demonstrate an improved learning of CzSL in comparison with original ones.

As somehow expected, the use of data augmentation has a positive impact in the performance for the majority of the tested models (but not always), including the two baselines (Expert and Amateur). However, the improvement of the baselines is generally limited w.r.t. what we may improve using the proposed CzSL. Looking at the results in detail, we see that such an improvement depends on the classification problem being addressed. For the binary and multi-class problems, both baselines improve slightly but still far from what CzSL may provide, especially if data augmentation is also considered within CzSL. In particular, the most remarkable improvement is achieved with the GZ1-(RLE) data set, where the implementation of data augmentation in Stages 2 and 3 of CzSL leads to a significant enhancement in the model's performance. In imbalanced classification problems, the results are more heterogeneous. When no pre-processing is applied, the performance is greatly reduced in all experiments, maybe due to the randomness of the data augmentation process that increases the difficulty derived from class imbalance, reinforcing the majority class. The results with RUS are also somewhat negative. However, it is remarkable the increase of performance obtained when ROS is considered. In this case, the use of data augmentation has resulted in a much better performance for the baselines, comparable to the one we initially obtained with CzSL. Nevertheless, CzSL also benefits substantially from using data augmentation in this setting. Thus, we consider that the use of data augmentation is an interesting avenue to improve further the results and we recommend to explore in conjunction with the proposed CzSL in the presence of low levels of labelled data.        

\begin{table*}[!h]
\centering
\resizebox{0.90\textwidth}{!}{
\begin{tabular}{lc||c||ccc}

\hline 
\textbf{Approach} & \textbf{GZ1-(SE)} & \textbf{GZ1-(RLE)} & \textbf{GZ1-(MG)} & \textbf{GZ1-(MG) + ROS} & \textbf{GZ1-(MG) + RUS} \\  
& Acc\textsubscript{\textit{TS}} & Acc\textsubscript{\textit{TS}} & G-mean\textsubscript{\textit{TS}} & G-mean\textsubscript{\textit{TS}} & G-mean\textsubscript{\textit{TS}} \\\hline

\textbf{\textit{Expert} (standard)} & \textbf{0.9225 $\pm$ 0.0047}  & \textbf{0.6367 $\pm$ 0.0071} & \textbf{0.2096 $\pm$ 0.0973} & 0.7834 $\pm$ 0.0091 & 0.6537 $\pm$ 0.0736 \\

\textbf{\textit{Expert} (augmented)} & 0.9207 $\pm$ 0.0280 & 0.6363 $\pm$ 0.0298 & 0.0743 $\pm$ 0.1028 & \textbf{0.8155 $\pm$ 0.236} & \textbf{0.6590 $\pm$ 0.0842} \\\hline

\textbf{\textit{Amateur} (standard)} & 0.9143 $\pm$ 0.0044 & 0.6468 $\pm$ 0.0055 & \textbf{0.2624 $\pm$ 0.1063} & 0.7482 $\pm$ 0.0066 & \textbf{0.8107 $\pm$ 0.0226} \\

\textbf{\textit{Amateur} (augmented)} & \textbf{0.9264 $\pm$ 0.0030} & \textbf{0.6773 $\pm$ 0.0353} & 0.1144 $\pm$ 0.0793 & \textbf{0.8147 $\pm$ 0.0336} & 0.7174 $\pm$ 0.0464 \\\hline

\textbf{\textit{CzSL} (standard)} & 0.9405 $\pm$ 0.0036 & 0.7192 $\pm$ 0.0052 & \textbf{0.7431 $\pm$ 0.0217} & 0.8144 $\pm$ 0.0049 & \textbf{0.8631 $\pm$ 0.0079} \\

\textbf{\textit{CzSL} (Stage 3 augmented)} & 0.9452 $\pm$ 0.0052 & 0.7167 $\pm$ 0.0171 & 0.6619 $\pm$ 0.0557 & 0.8470 $\pm$ 0.0140 & 0.7999 $\pm$ 0.0383 \\

\textbf{\textit{CzSL} (Stages 2 and 3 augmented)} & \textbf{0.9468 $\pm$ 0.0022} & \textbf{0.7409 $\pm$ 0.0254} & 0.7052 $\pm$ 0.0615 & \textbf{0.8501 $\pm$ 0.0331} & 0.8270 $\pm$ 0.0107 \\\hline

\end{tabular}}
\caption{\label{tab:data-augmentation} Results of the data augmentation study with Expert and Amateur comparative approaches, and the CzSL model. The best result is highlighted for each approach.}
\end{table*}

\subsection{Ablation study of CzSL}
\label{ablation-study}
\begin{figure}
\centering 
\resizebox{1.0\columnwidth}{!}{\includegraphics{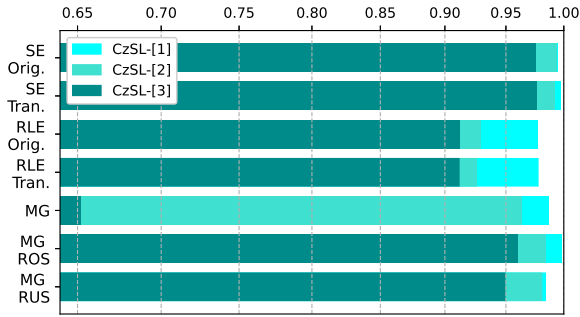}} 
\caption{Comparative view of CzSL ablated versions' performance. CzSL's performance corresponds to 1.0. The number between square brackets refers to the learning stage that is ablated (Figure \ref{fig:CzSL-workflow}).}
\label{fig:ablation-study}
\end{figure}

Finally, we conduct an ablation study to quantify separately the contribution of the learning stages of the CzSL algorithm, which leverage the unlabelled, amateur- and expert-labelled data, respectively (Figure \ref{fig:CzSL-workflow}). We compute the performance of an altered (ablated) version of CzSL, eliminating one of the three stages each time. We refer to these approaches as CzSL-[1], CzSL-[2], and CzSL-[3], indicating between square brackets the stage that is ablated in accordance with the description of the learning workflow (Section \ref{citizen-science-learning}). These results are shown in Table \ref{tab:ablation-study} for the three data sets investigated, and original and transformed scores. For simplicity, we restrict the comparison to performance metrics on $ TS $ data. We also present in Figure \ref{fig:ablation-study} a comparison of the ablated versions' performances versus the (standard) CzSL's performance.

\begin{table*}[!ht]
\centering
\resizebox{1.00\textwidth}{!}{
\begin{tabular}{lcc||cc||ccc}

\hline 
\textbf{Approach} & \multicolumn{2}{c}{\textbf{GZ1-(SE)}} & \multicolumn{2}{c}{\textbf{GZ1-(RLE)}} & \textbf{GZ1-(MG)} & \textbf{GZ1-(MG) + ROS} & \textbf{GZ1-(MG) + RUS} \\ 
& \textbf{\textit{Original}} & \textbf{\textit{Transformed}} & \textbf{\textit{Original}} & \textbf{\textit{Transformed}} & & & \\ 
& Acc\textsubscript{\textit{TS}} & Acc\textsubscript{\textit{TS}} & Acc\textsubscript{\textit{TS}} & Acc\textsubscript{\textit{TS}} & G-mean\textsubscript{\textit{TS}} & G-mean\textsubscript{\textit{TS}} & G-mean\textsubscript{\textit{TS}} \\\hline

\textbf{\textit{CzSL} (standard)} & 0.9372 $\pm$ 0.0029 & 0.9405 $\pm$ 0.0036 & 0.7167 $\pm$ 0.0061 & 0.7192 $\pm$ 0.0052 & 0.7431 $\pm$ 0.0217 & 0.8144 $\pm$ 0.0049 & 0.8631 $\pm$ 0.0079 \\\hline 

\textbf{\textit{CzSL-[1]}} & 0.9325 $\pm$ 0.0042 & 0.9378 $\pm$ 0.0019 & 0.7002 $\pm$ 0.0052 & 0.7033 $\pm$ 0.0029 & 0.7331 $\pm$ 0.0219 & 0.8130 $\pm$ 0.0020 & 0.8496 $\pm$ 0.0375 \\

\textbf{\textit{CzSL-[2]}} & 0.9313 $\pm$ 0.0026 & 0.9327 $\pm$ 0.0016 & 0.6658 $\pm$ 0.0055 & 0.6660 $\pm$ 0.0026 & 0.7162 $\pm$ 0.0260 & 0.8018 $\pm$ 0.0037 & 0.8468 $\pm$ 0.0186 \\

\textbf{\textit{CzSL-[3]}} & 0.9144 $\pm$ 0.0059 & 0.9180 $\pm$ 0.0017 & 0.6535 $\pm$ 0.0074 & 0.6558 $\pm$ 0.0044 & 0.4845 $\pm$ 0.0413 & 0.7418 $\pm$ 0.0129 & 0.8197 $\pm$ 0.0160 \\\hline

\end{tabular}}
\caption{\label{tab:ablation-study} Results of the CzSL ablation study. The number between square brackets refers to the learning stage that is ablated.}
\end{table*}

On the basis of these results, we elucidate that the three stages of CzSL contribute to the overall performance investigated in previous sections. However, they do not equally weigh in the algorithm's learning. The ablation study demonstrates that the learning increases in the bottom-up direction, that is to say, the results for ablated versions progressively decrease from Stage 1 to Stages 2 and 3. This indicates that the learning is more significant as higher levels of knowledge are exploited, from unlabelled data to amateur and expert labels. This fact is in accordance with the assumption made as part of the motivation to exploit the three levels of knowledge available, i.e. small amounts of data are worth considering whether they hold a high quality labelling. Moreover, this trend shows clearer as the classification problem entails more difficulty: there are tiny discrepancies between the ablated versions in the binary problem, but these get more pronunciated for the imbalanced problem and the multi-class problem, as the one presenting the greatest differences. Particularly, the CNN pre-training and fine-tuning phases prove to be key in the experiments with GZ1-(RLE) data set, for which the accuracy decreases around 5-6\% if we eliminate any of the two stages of the CzSL learning. Additionally, the ablation of the fine-tuning phase drastically reduces the performance for the imbalanced problem when no pre-processing is applied.

\section{Conclusions}
\label{conclusions} 
In this paper, we have presented Citizen Science Learning, CzSL, a novel learning methodology especially devised to exploit the labelling framework deployed around the development of citizen science projects on the web towards an improved image classification in astronomy. CzSL makes it possible to learn from data labelled by professional and amateur astronomers, and from unlabelled data, representing the three levels of knowledge available and worthy of consideration in modern citizen science. It is a deep learning-based methodology based on two key elements: a well-established form of transfer learning, the pre-training and fine-tuning of convolutional neural networks, and the extension of expert knowledge to citizen science data. CzSL first learns patterns from unlabelled data via the pre-training of the network with a convolutional autoencoder, and then exploits the limited expert knowledge to improve the accuracy of labels that are derived from a set of amateur classifications, featuring a transfer of labels between experts and amateurs. In summary, the CzSL algorithm has demonstrated a more integrated use of citizen science data and classifications validated by professional astronomers that greatly enlarges both citizen science and expert efforts utility towards more robust and reliable automated classifiers. 

As future work, we will consider a thorough exploration of more advanced semi-supervised and self-supervised strategies to further improve the results and competitiveness of the proposed method compared to current approaches that exploit labelled and unlabelled data. As well, an in-depth investigation of data augmentation techniques for the classification problem at hand towards a more efficient use of expert-labelled data, the most difficult data to obtain, might entail several modifications to the deep learning solutions and architectures explored in this research. However, these upgrades could also provide fruitful insights at the boundaries between the use of high-quality labelled data, unlabelled data, and citizen science.

\section*{Acknowledgements}
Authors Manuel Jim\'enez and Emilio J. Alfaro acknowledge financial support from the State Agency for Research of the Spanish MCIU through the ``Center of Excellence Severo Ochoa'' award to the Instituto de Astrof\'isica de Andaluc\'ia (SEV-2017-0709). This work is supported by projects A-TIC-434-UGR20 and PID2020-119478GB-I00. We gratefully acknowledge the support of NVIDIA Corporation with the donation of the Titan Xp GPU used in this research. We also thank Dr. Steven Bamford (University of Nottingham) for the valuable discussions about the \textit{Galaxy Zoo} project and data.    

\section*{Data Availability}
Data and Python code underlying this research are available at \url{https://github.com/manueljimenez86/CzSL}, including the original GZ1 results and galaxy classifications from both volunteers and CzSL algorithm as well as the images used in the CzSL training.




\bibliographystyle{mnras}
\bibliography{MN-22-0500-MJ.R4-ref} 


\appendix
\section{Data transformations}
\label{appendix-A}
The approach presented in \citet{Jimenez2019479} aims to leverage the uncertainty within data labelled by amateur participants in CzS projects, leading to improved classifications and a better exploitation of CzS data. Such an approach assumes a \textit{Don't Know} (DK) option, that is, an alternative to be clicked on by participants whether the image is blurry, ambiguous, or difficult to be assigned to any other of the classes, and the availability of expert classifications, which are taken as ground truth. The method is composed of two phases. First, a set of three mathematical transformations is applied to the CzS scores. A hybridisation strategy then explores the best joint combination of these based on their evaluation against the ratings of experts. 

Let $ X = (x_1, x_2,..., x_C) $, $ x_i = \frac{\nu_i}{N} $, be the score vector for a given example, with $ C $ the number of options offered to participants in the web, $ \nu_i $ the number of votes assigned to option $ i $, and $ N $ the total number of votes received by the example. The transformations are defined as follows:
\begin{itemize}

\item \textbf{Normalisation \{1\}}: the scores are normalised according to the main classes of the problem. These are the classes holding a greater importance with respect to the rest and that represent the target of the classification problem being addressed. For instance, in GZ1 we highlight the \textit{Elliptical} and \textit{Spiral} scores against the rest, \textit{Don't Know} and \textit{Merger}. The normalised score vector $ Z = (z_1, z_2,..., z_M) $ is obtained by computing a normalisation $ z_i = \frac{x_i}{\sum x_i} $, for $ i \in \lbrace 1,2...M \rbrace $ and $ M $ the number of main classes.

\item \textbf{DK votes shift \{2\}}: the main classes' scores $ \widehat{X} = (x_1, x_2,..., x_M) $ are modified using the information about the uncertainty held in DK votes. The shifted score vector $ \widehat{W} = (w_1, w_2,..., w_M) $ is calculated first computing $ \epsilon = \frac{\alpha \cdot \mu_{DK}}{\beta + \nu_{DK}} $ for the example, and then applying $ w_{i} = x_{i} + \epsilon $  to the favoured class, or $ w_{i} = x_{i} - \frac{\epsilon}{M - 1} $ in other case, with $ \mu_{DK} $ the average number of DK votes across the entire data set, $ \nu_{DK} $ the number of DK votes for the example, and $ \alpha $ and $ \beta $ parameters that are adjusted by testing a set of pair of values and evaluating the modified scores with expert classifications. The \textit{favoured} class of the problem is the class that is prone to receive more clicks whenever there is not evidence enough to assign a different one. In GZ1, this is the case for \textit{Elliptical}, as in the absence of a good resolution, an image of a spiral galaxy will tend to be classified as elliptical. 

\item \textbf{Votes boost \{3\}}: each score is modified using the distribution for that class across the entire data. The boosted score vector $ \widehat{R} = (r_1, r_2,...,r_M) $ is defined as $ r_i = x_i + \sigma \mathrm{sigmoid}(\tilde{v}_{i}) $, with $ \tilde{v}_{i} = \frac{\nu_i - \mu_i}{\sigma_i} $, $ \mu_i $ and $ \sigma_i $ the mean and standard deviation over the entire data set for the class, $ \mathrm{sigmoid}(x) = \frac{1}{1 + e^{-x}} $ the sigmoid function, and $ \gamma \in \left[0, 1 \right]  $ a parameter to be adjusted similarly to $ \alpha $ and $ \beta $ parameters introduced above. 

\end{itemize}
Once the three transformations are computed, an hybridisation strategy explores the best concatenation of them, that is, the order in which they are applied to the original data. For this, the transformations are computed following all possible combinations. This yields a set of $ \binom{3}{1} + \binom{3}{2}\cdot2! + \binom{3}{3}\cdot3! = 15 $ transformation sequences, which are evaluated using expert classifications as ground truth. Finally, we obtain a ranking of sequences \{\textit{xyz}\} that informs about the most adequate application of the transformations described above.

In the experiments carried out to test the CzSL algorithm, we employ the \{123\} transformation sequence, which applies the \textbf{Normalisation}, \textbf{DK votes shift}, and \textbf{Votes boost} in this order to the original GZ1 scores. This sequence demonstrated the best performance in the case study conducted with the GZ1 data. For a complete description of the method and the study, we refer to \citet{Jimenez2019479}.  

\section{MLP design}
\label{appendix-B} 
In order to select the best MLP architecture through experimentation, we conduct a grid search to find the model with the best overall performance for the experiments with the classification problems covered. We first establish a set of architectures involving a varied number of layers and neurons. Then, a nested cross-validation \citep{Varma2006} is carried out considering the $ E $ set, which is split in 80/20 to complete a five-fold cross-validation. This way, the examined MLP architectures are first trained on the 80\% of $ E $ and then ranked by evaluating the performance of each model on the test data (20\%), comparing the MLP predictions with the actual expert label. The obtained ranking for the three data sets explored is in Table \ref{tab:MLP-model-nest-val}, where the number of hidden layers and neurons of the tested models are indicated in brackets (e.g. the [7,5,3] model consists of three hidden layers holding 7, 5, and 3 neurons, respectively). For binary and multi-class classification problems, the values indicated correspond to the average Acc over the five test data partitions. For the imbalanced classification problem, to the average G-mean.

\begin{table}[!h]
\centering
\resizebox{0.75\columnwidth}{!}{
\begin{tabular}{lccc}

\hline
\textbf{Architecture} & \textbf{Binary} & \textbf{Multi-class} & \textbf{Imbalanced} \\
 & (Acc\textsubscript{\textit{TS}}) & (Acc\textsubscript{\textit{TS}}) & (G-mean\textsubscript{\textit{TS}}) \\\hline
\textbf{[8,7,5,3]}    & \textbf{0.9584} & \textbf{0.9419} 	   & \textbf{0.9773} \\ 
\textbf{[7,5,4,3]}    & 0.9578          & 0.5159  			   & 0.3911 \\ 
\textbf{[7,5,3]}      & \textbf{0.9584} & 0.9328  			   & 0.9617 \\
\textbf{[5,3]}        & 0.9462          & 0.9352  			   & 0.9663 \\
\textbf{[3]}          & 0.9266          & 0.5434  			   & 0.0000 \\\hline
								   	
\end{tabular}}
\caption{\label{tab:MLP-model-nest-val} Results for the MLP architectures testing implementing 80/20 nested cross-validation over the $ E $ data set. The best result within each classification problem has been highlighted in bold.}
\end{table}

An evaluation of the obtained results reveals that the best-performing MLP architecture is composed of four hidden layers with 8, 7, 5, and 3 neurons. As with the CNN implemented (Table \ref{tab:CNN-CAE}), SoftMax is applied to the last layer to provide class probabilities.

\bsp	
\label{lastpage}
\end{document}